\documentclass[journal,twocolumn,compress]{IEEEtran}

\usepackage{dsfont}
\usepackage{moreverb}
\usepackage{epsfig}
\usepackage{amsmath,amssymb,amsthm,mathrsfs,amsfonts,dsfont,bm}
\usepackage{algorithm}
\usepackage{algorithmic}
\usepackage{subfigure}
\usepackage{multirow}
\usepackage{rotating}
\usepackage{tabularx}
\usepackage{array}
\usepackage{anyfontsize}
\usepackage{color,soul}
\usepackage{graphicx,dblfloatfix}
\usepackage{epstopdf}
\usepackage{blindtext}
\usepackage{fancyhdr}
\usepackage{tcolorbox}
\usepackage{setspace}
\usepackage{enumitem}
\usepackage{diagbox} % for diagonal cell in table
\usepackage{tikz}
\usetikzlibrary{patterns}
\usetikzlibrary{patterns.meta}
\usepackage{pgfplots}
\pgfplotsset{compat=1.18}
\tikzset{
  labellargefont/.style={font=\large\color{white!15!black}},
  labelLargefont/.style={font=\Large\color{white!15!black}},
  labelLARGEfont/.style={font=\LARGE\color{white!15!black}},
  labelhugefont/.style={font=\huge\color{white!15!black}}
}

\usetikzlibrary{arrows.meta}
\usetikzlibrary{plotmarks}
\usepackage{enumitem}
\def\tran{\text{T}}
\def\Com{\mathrm{Com}}
\def\Rad{\mathrm{Rad}}

\def\H{\mathrm{H}}
\def\vec{\mathrm{vec}}
\def\diag{\mathrm{diag}}

\def\Real{\Re}

\newcommand{\datafolder}{FigureDataBistatic/}

\allowdisplaybreaks

\usepackage{pifont}

\usepackage{colortbl}

\usepackage{booktabs}   % \toprule, \midrule, \bottomrule
\usepackage{makecell}   % for multi-line header cells
\usepackage{xcolor}     % colors
\graphicspath{{Figure/}}

\usepackage{cite}

\begin{document}

\title{Beamforming and Filter Design for Bistatic ISAC under Known and Unknown Transmit Symbols}

% Joint Beamforming and Radar Filter Design for Bistatic Multiuser MIMO ISAC: Effects of Known versus Unknown Transmit Symbols

\author{Mohammad Hatami, 
Nhan Thanh Nguyen, 
and
Markku Juntti, 
\thanks{
Mohammad Hatami, Nhan Thanh Nguyen, and Markku Juntti are with the Center for Wireless Communications, University of Oulu, FI-90014 Oulu, Finland (e-mail:
mohammad.hatami@oulu.fi; nhan.nguyen@oulu.fi; markku.juntti@oulu.fi).\\
% This work has been financially supported by Infotech Oulu, 6G-WISECOM (grant number: 24304960), 6G-ISAC (grant number: 24304492), the Research Council of Finland through 6G Flagship Programme (grant number: 346208) and through project DIRECTION (grant number: 354901), and CHIST-ERA via project PASSIONATE (grant number: 359817).
}
}

% \author{
% \IEEEauthorblockN{
%     Mohammad Hatami, Nhan Thanh Nguyen, and Markku Juntti
% 	\\}
% \IEEEauthorblockA{
% 	\small%
% 	  Centre for Wireless Communication, University of Oulu, 90014 Oulu, Finland \\
%       e-mails: mohammad.hatami@oulu.fi; nhan.nguyen@oulu.fi; markku.juntti@oulu.fi
% }
% }
% make the title area
\maketitle

\begin{abstract}
This paper investigates the joint design of beamforming and radar receive filters in a multiuser bistatic integrated sensing and communications (ISAC) system, aiming to maximize the minimum radar signal-to-interference-plus-noise ratio (SINR) under communications SINR and transmit power constraints.
We consider two scenarios: transmitted signals are either known or unknown at the radar receiver. We develop tractable solutions to the resulting non-convex optimization problems in both cases. For the known-signal case, we derive closed-form radar receive filters and iteratively design beamforming using fractional programming (FP) and successive convex approximation (SCA). For the unknown case, we adopt an alternating optimization (AO) approach to jointly design the beamforming and receive filters.  
Numerical results demonstrate that, while both approaches achieve comparable performance under per-slot optimization, knowledge of the transmitted symbols provides significant gains in multi-slot processing via coherent integration. Moreover, the proposed ISAC designs perform close to the radar-only benchmark under moderate communication requirements.
\end{abstract}
\vspace{-4mm}
\vspace{-3mm}
\section{Introduction}
Integrated sensing and communications (ISAC) is an emerging technology for next-generation wireless networks due to its advantages in resource efficiency, mutual performance gains, and its potential for spectrum and hardware sharing \cite{zhang2021overview,dong2025ComAssistedSensing}.
% zhou2022integrated, dong2025ComAssistedSensing, Mandelli2023SurveyISAC
ISAC enables the simultaneous operation of sensing and communication functionalities using a common signal waveform, making transmit waveform and beamforming design a critical aspect of overall system performance.
Moreover, in bistatic systems, the presence of radar receive (Rx) filters introduces additional design variables. This motivates the joint design of transmit beamforming and Rx filters under communications and power constraints, which is the focus of this work.

Recent research has extensively investigated transmit beamforming and waveform design for ISAC systems.
To evaluate sensing performance, a variety of metrics have been adopted in the literature, including beampattern-based measures \cite{liu2018toward, hua2023optimal,Nhan2024JCASDeep,wu2025quantized,Liu2023JointTxRxFullDuplex}, spatial spectrum matching error (SSME) \cite{cheng2021HBF-OFDM}, the Cramér–Rao lower bound (CRLB) \cite{He2026BistaticEnhancement,Mao2025HBFBistatic, ren2022fundamental, nguyen2024massive,Zhong2025Resource}, mutual information (MI) \cite{wei2024waveform}, signal-to-clutter-plus-noise ratio (SCNR) \cite{choi2024joint, wang2025lowcomplexity}, and signal-to-interference-plus-noise ratio (SINR) \cite{Wen2023EffTransISAC, Zhao2025JointBF,johnston2022mimo, Liu2024JointTxRxMultiStatic, jiang2025fullduplex,Hatami2026BF_TCOM,Hatami2024WaveformDesignSPAWC,Hatami2024JointWaveformDesignJCS,hatami2025EEwcnc}. In this work, we adopt SINR as the primary sensing performance metric.
Most existing works focus on monostatic radar systems, where the transmitter and receiver are co-located. Such systems typically operate in full-duplex mode and require interference cancellation techniques to mitigate self-interference from the transmitter to the receiver. However, this aspect is often overlooked in the literature. As an alternative, bistatic radar systems, where the transmitter and receiver are geographically separated, provide a practical solution by inherently avoiding strong self-interference. Motivated by this, we consider a bistatic ISAC system in this paper.

We consider two scenarios based on the availability of transmitted symbols at the radar receiver. This distinction is important in practice, as symbol availability depends on coordination between the transmitter and sensing receiver. In some systems, symbols can be shared and exploited for improved sensing, while in others, they may be unavailable due to system constraints or privacy considerations.
In the first scenario, the symbols are shared, allowing the receiver to exploit this knowledge in designing the receive filters. In the second, the symbols are unavailable. In both cases, we jointly design the transmit beamforming and radar receive filters. Numerical results show that, for a frame length of one (per-slot optimization), performance without symbol knowledge is close to the known-symbol case. However, when the frame length exceeds one symbol, multiple observations enable spatio-temporal adaptive processing (STAP), where symbol knowledge allows coherent integration, while unknown symbols yield unstructured observations, limiting multi-slot gains.

\section{System Model and Problem Formulation}\label{sec:systemmodel_bistatic}
We consider a bistatic MIMO ISAC system. The base station (BS) is equipped with $N$ transmit antennas and serves $U$ single-antenna downlink users while simultaneously sensing $Q$  targets. In addition, a spatially separated sensing receiver (Rx) equipped with $M$ antennas collects the reflected signals.
% \red{We consider a bistatic multiuser MIMO ISAC system, where the base station (BS) transmits probing signals towards $Q$ sensing targets while simultaneously providing downlink communication services to $U$ single-antenna users. In addition, a spatially separated sensing receiver (Rx) is deployed to collect the reflected signals. The BS is equipped with $N$ transmit (Tx) antennas, while the sensing receiver is equipped with $M$ receive (Rx) antennas.}

\subsection{Communications Model}
Let $\mathbf{h}_u \in \mathbb{C}^{N \times 1}$ denote the channel vector from the BS to user $u$, and $\mathbf{f}_u \in \mathbb{C}^{N \times 1}$ denote the corresponding beamforming vector. The channels are assumed perfectly known (see e.g., \cite{liu2018toward, hua2023optimal,Nhan2024JCASDeep,wu2025quantized,Liu2023JointTxRxFullDuplex,cheng2021HBF-OFDM,He2026BistaticEnhancement,Mao2025HBFBistatic, ren2022fundamental}) at the transmitter and remain constant over a transmission frame of $L$ time slots (see e.g., \cite{liu2018toward,wu2025quantized,ren2022fundamental,wang2025lowcomplexity,Wen2023EffTransISAC}).
Let $s_u[l]$ denote the transmitted symbol for user $u$ at time slot $l$, with average unit power $\mathbb{E}\{|s_u[l]|^2\} = 1$.
Defining the data vector  $\mathbf{s}[l] = [s_{1}[l], \dots ,s_{U}[l]] \in \mathbb{C}^{U\times1}$ and beamforming matrix $\mathbf{F} = [\mathbf{f}_{1},\dots, \mathbf{f}_{U}] \in \mathbb{C}^{N\times U}$, the transmitted signal at slot $l$ is  $\mathbf{x}[l] = \mathbf{F} \mathbf{s}[l]$. We define the signal matrix over $L$ slots as 
$\mathbf{X} = [\mathbf{x}[1],\dots,\mathbf{x}[L]] \in \mathbb{C}^{N\times L}$.
The received signal at user $u$ at time slot $l$ is given by
\begin{align}
    y_{u}&[l]  = \mathbf{h}_{u}^{\H} \mathbf{x}[l] +{n}_{u}[l] = \mathbf{h}_{u}^{\H} \mathbf{F} \mathbf{s}[l] +{n}_{u}[l]\notag\\ 
    & = \mathbf{h}_{u}^{\H} \mathbf{f}_{u} s_{u}[l] + {\sum_{j\neq u}}\mathbf{h}_{u}^{\H} \mathbf{f}_{j} s_{j}[l] + {n}_{u}[l],
\end{align}
where $n_u[l] \sim \mathcal{CN}(0,\sigma_C^2)$.
% \red{where $n_{u}[l]$ denotes additive white Gaussian noise (AWGN) with zero mean and variance $\sigma^2_\C$, i.e., $n_{u}[l] \sim \mathcal{CN}(0,\sigma^2_\C)$.}
Accordingly, the average SINR for user $u$ is expressed as
\begin{align}\label{eq:SINR_Com}
    \mathrm{SINR}^{\Com}_{u} 
    % & = \frac{1}{L}\sum_{l = 1}^{L}\frac{|\mathbf{h}_{u}^{\H} \mathbf{f}_{u}s_{u}[l]|^2}{\sum_{j = 1, j\neq u}^{U} |\mathbf{h}_{u}^{\H} \mathbf{f}_{j}s_{j}[l]|^2 + \sigma^2_\C}\\
    & = \frac{1}{L}\sum_{l = 1}^{L} \frac{|\tilde{\mathbf{h}}_{u}^{\H} \mathbf{f}_{u}s_{u}[l]|^2}{\sum_{j = 1, j\neq u}^{U} |\tilde{\mathbf{h}}_{u}^{\H} \mathbf{f}_{j}s_{j}[l]|^2 + 1}.
\end{align}
where $\tilde{\mathbf{h}}_{u} \triangleq \tfrac{\mathbf{h}_{u}}{\sigma_{u}}$ is the normalized channel coefficients with respect to the noise standard deviation. Averaging over the transmit symbols, we obtain
\begin{align}\label{eq:SINR_Com_Avg}
    \overline{\mathrm{SINR}}^{\Com}_{u} = \frac{|\tilde{\mathbf{h}}_{u}^{\H} \mathbf{f}_{u}|^2}{\sum_{j = 1, j\neq u}^{U} |\tilde{\mathbf{h}}_{u}^{\H} \mathbf{f}_{j}|^2 + 1}.
\end{align}

% We define the average SINR at user $u$ over communications subcarriers in set $\mathcal{G}$  as
% \begin{align}\label{eq:SINR_Com}
%     \mathrm{SINR}^{\Com}_{u} & = \frac{1}{|\mathcal{G}|}\sum_{k \in \mathcal{G}} \frac{|\mathbf{h}_{u,k}^{\H} \mathbf{f}_{u,k}|^2}{\sum_{j = 1, j\neq u}^{U} |\mathbf{h}_{u,k}^{\H} \mathbf{f}_{j,k}|^2 + \sigma^2_\C}\\
%     & = \frac{1}{|\mathcal{G}|}\sum_{k \in \mathcal{G}} \frac{|\tilde{\mathbf{h}}_{u,k}^{\H} \mathbf{f}_{u,k}|^2}{\sum_{j = 1, j\neq u}^{U} |\tilde{\mathbf{h}}_{u,k}^{\H} \mathbf{f}_{j,k}|^2 + 1}, 
% \end{align}
% where $\tilde{\mathbf{h}}_{u,k} \triangleq \tfrac{\mathbf{h}_{u,k}}{\sigma_{u,k}}$ is the normalized channel coefficients with respect to the noise standard deviation.

\subsection{Sensing/Radar Model}
In bistatic setup, the $Q$ targets are characterized by their angle of departure (AoD) $\{\phi_1,\ldots,\phi_Q\}$, angle of arrival (AoA) $\{\theta_1,\ldots,\theta_Q\}$, and propagation delays. Let $\{l_0, l_1, \ldots, l_Q\}$ denote the range bins\footnote{The range bin is calculated as 
$l = \left\lfloor \frac{\tau}{T_s} \right\rfloor 
= \left\lfloor \frac{d_{\mathrm{Tx}\to q} + d_{q \to \mathrm{Rx}}}{c\,T_s} \right\rfloor,$
where $\tau$ denotes the propagation delay, $T_s$ is the sampling period, $c$ is the speed of light, and $d_{\mathrm{Tx}\to q}$ and $d_{q \to \mathrm{Rx}}$ represent the distances from the transmitter to target $q$ and from the target to the receiver, respectively. 
For the LOS path, the range bin is given by
$l_0 = \left\lfloor \frac{d_{\mathrm{Tx}\to \mathrm{Rx}}}{c\,T_s} \right\rfloor,$
where $d_{\mathrm{Tx}\to \mathrm{Rx}}$ denotes the direct distance between the transmitter and the receiver.} corresponding to the line-of-sight (LOS) path and the $Q$ targets, respectively, where $l_0 \leq l_1 \leq \dots \leq l_Q$. 
The bistatic response matrix for target $q$ is defined as
\begin{align}
    \mathbf{A}_{q} = \mathbf{a}^{\text{Rx}}(\theta_q) (\mathbf{a}^{\text{Tx}}(\phi_q))^{\mathrm{T}} \in \mathbb{C}^{M\times N},
\end{align}
% \begin{align}
%     \mathbf{A}_{q} = \mathbf{a}_{q}^{\text{Rx}} (\mathbf{a}_{q}^{\text{Tx}})^{\mathrm{H}} \in \mathbb{C}^{M\times N},
% \end{align}
where $\mathbf{a}^{\text{Rx}}(\theta_q)$ and $\mathbf{a}^{\text{Tx}}(\phi_q)$ denote the steering vectors at the Rx and Tx arrays, respectively.
Assuming uniform linear arrays (ULAs), these vectors are given by
\begin{align}
    & \mathbf{a}^{\text{Tx}}(\phi_q) = \frac{1}{\sqrt{N}} \left[1,\dots,e^{-j2\pi(N-1)d\sin(\theta_q^{\text{Rx}})/\lambda}\right]^{\tran},\\
    & \mathbf{a}^{\text{Rx}}(\theta_q) = \frac{1}{\sqrt{M}} \left[1,\dots,e^{-j2\pi(M-1)d\sin(\theta_q^{\text{Tx}})/\lambda}\right]^{\tran},
\end{align}
where $d$ denotes the antenna spacing and $\lambda$ is the carrier wavelength. 
In addition to reflections from the targets, the Rx also observes a direct LOS component.
Accordingly, the received signal matrix is expressed as
\begin{align}
\label{eq_radar_model_bistatic}
    \mathbf{Y} = 
    \sum_{q = 0}^{Q}\beta_{q} \mathbf{A}_{q} \tilde{\mathbf{X}} \mathbf{J}_{l_q-l_0}
    + \mathbf{N},
\end{align}
% \begin{align}
% \label{eq_radar_model_bistatic}
%     \mathbf{Y} & =
%     \beta_{0} \mathbf{A}_{0} \tilde{\mathbf{X}}
%     +
%     \sum_{q = 1}^{Q}\beta_{q} \mathbf{A}_{q} \tilde{\mathbf{X}} \mathbf{J}_{l_q-l_0}
%     + \mathbf{N}\\
%     & = 
%     \sum_{q = 0}^{Q}\beta_{q} \mathbf{A}_{q} \tilde{\mathbf{X}} \mathbf{J}_{l_q-l_0}
%     + \mathbf{N},
% \end{align}
where $\beta_{0}$ and $\mathbf{A}_{0}$ denote the complex channel gain and spatial response of the LOS path, respectively. The parameter $\beta_{q}$ represents the complex gain of target $q$, capturing the bistatic propagation loss and radar cross section (RCS). The matrix $\tilde{\mathbf{X}} = [\mathbf{X}, \mathbf{0}_{N\times(l_Q-l_0)}] \in \mathbb{C}^{N\times (L+l_Q-l_0)}$ is the zero-padded transmit waveform matrix, and $\mathbf{N} \in \mathbb{C}^{M\times (L+l_Q-l_0)}$ denotes additive white Gaussian noise with entries distributed as $\mathcal{CN}(0,\sigma^2_\mathrm{R})$. The matrix $\mathbf{J}_{l_q - l_0}$ denotes the delay operator relative to the LOS path, with entries given by
\begin{equation}
    \mathbf{J}_{l}(i,j) = 
    \begin{cases}
        1 & j - i = l, \\
        0 & \text{otherwise}.
    \end{cases}
\end{equation}

Let $\mathbf{w}_{q} \in \mathbb{C}^{M(L+l_Q-l_0)\times 1}$ denote the space-time receive filter for detecting target $q$. The filter output is given by
\begin{equation}\label{eq:out_filter_radar_bistatic}
    y_q = \mathbf{w}_{q}^{\mathrm{H}}\vec(\mathbf{Y}) = \beta_{q} \mathbf{w}_{q}^{\mathrm{H}}\tilde{\mathbf{A}}_{q}\mathbf{x} + \sum_{\bar{q} \neq q} \beta_{\bar{q}} \mathbf{w}_{q}^{\mathrm{H}}\tilde{\mathbf{A}}_{\bar{q}}\mathbf{x} + n_{q},\nonumber
\end{equation}
where $\mathbf{x} = \vec(\mathbf{X})$, $\tilde{\mathbf{A}}_{q}= (\mathbf{J}_{l_0-l_q} \otimes \mathbf{A}_{q})\mathbf{T}$, $\mathbf{T} = [\mathbf{I}_{N L},\mathbf{0}_{NL \times N(l_Q-l_0)}]^{\tran}$, and $n_{q} = \mathbf{w}_{q}^{\mathrm{H}} \vec({\mathbf{N}})$. 
By defining $\tilde{\mathbf{F}} = \diag(\mathbf{F}, \dots,\mathbf{F}) = \mathbf{I}_{L} \otimes \mathbf{F} \in \mathbb{C}^{NL\times UL}$ and $\mathbf{s} =  [\mathbf{s}^\tran[1],\dots,\mathbf{s}^\tran[L]]^\tran \in \mathbb{C}^{UL\times 1}$, we have  $\mathbf{x} = \tilde{\mathbf{F}} \mathbf{s}$.

We define the sensing SINR for target $q$ as the ratio of the desired echo power to the aggregate interference, which includes echoes from other targets, the LOS component, and noise. Accordingly, the sensing SINR for target $q$ is given by
\begin{align}\label{eq:SINR_radar}
    \mathrm{SINR}^{\Rad}_q = \frac{|\beta_{q}|^2|\mathbf{w}_{q}^{\mathrm{H}}\tilde{\mathbf{A}}_{q}\mathbf{x}|^2}{ \mathbf{w}_{q}^{\mathrm{H}}\mathbf{R}_{q}\mathbf{w}_{q}}
\end{align}
where the interference-plus-noise covariance matrix is
\begin{align}
    \mathbf{R}_{q} = \sum_{\bar{q} \neq q} |\beta_{\bar{q}}|^2 \tilde{\mathbf{A}}_{\bar{q}}\mathbf{x}\mathbf{x}^{\mathrm{H}} \tilde{\mathbf{A}}_{\bar{q}}^{\mathrm{H}} + \sigma^2_R \mathbf{I}.
\end{align}

If the transmitted symbols $\mathbf{s}$ are unknown at the radar receiver, the sensing SINR is defined as
\begin{align}\label{eq:SINR_radar_avg}
    \hspace{-2mm}\overline{\mathrm{SINR}}^{\Rad}_q 
    \!\!\!= \!\!\frac{\mathbb{E}_{\mathbf{s}}\{|\beta_{q}|^2|\!\mathbf{w}_{q}^{\mathrm{H}}\tilde{\mathbf{A}}_{q}\mathbf{x}|^2\!\}}
    {\mathbb{E}_{\mathbf{s}}\{\mathbf{w}_{q}^{\mathrm{H}}\mathbf{R}_{q}\mathbf{w}_{q}\}} 
    % \notag \\&
    \!\!=\!\! \frac{|\beta_q|^2 \!\mathbf{w}_{q}^{\mathrm{H}}\!\tilde{\mathbf{A}}_{q}\tilde{\mathbf{F}}\tilde{\mathbf{F}}^{\mathrm{H}}\!\tilde{\mathbf{A}}_{q}^{\mathrm{H}}\!\mathbf{w}_{q}}
    {\mathbf{w}_{q}^{\mathrm{H}}\bar{\mathbf{R}}_{q}\mathbf{w}_{q}},
\end{align}
where we have used $\mathbb{E}\{\mathbf{s}\mathbf{s}^{\mathrm{H}}\} = \mathbf{I}$ and $\bar{\mathbf{R}}_{q}$ is given by
\begin{align}
    \bar{\mathbf{R}}_{q} & = \mathbb{E}_{\mathbf{s}}\{\mathbf{R}_{q}\}
    = \sum_{\bar{q} \neq q} |\beta_{\bar{q}}|^2 
    \tilde{\mathbf{A}}_{\bar{q}}\tilde{\mathbf{F}}\tilde{\mathbf{F}}^{\mathrm{H}} \tilde{\mathbf{A}}_{\bar{q}}^{\mathrm{H}}
    % + |\beta_{0}|^2 \tilde{\mathbf{A}}_{0}\tilde{\mathbf{F}}\tilde{\mathbf{F}}^{\mathrm{H}} \tilde{\mathbf{A}}_{0}^{\mathrm{H}}
    + \sigma^2_{\mathrm{R}} \mathbf{I}.
\end{align}

\subsection{Problem Formulation}
The sensing SINR depends on the beamforming $\mathbf{F}$ and receive filters $\mathbf{w}_{q}$. We therefore optimize $\mathbf{F}$ and $\mathbf{w}_{q}$, $\forall q$, to maximize the minimum radar SINR subject to per-user communications SINR and transmit power constraints. The optimization problem is then formulated as:
\begin{subequations}\label{eq:ProblemFormulationMain}
\begin{align}
    (\mathbf{P1}) ~~~&  \max_{\mathbf{F},\{\mathbf{w}_{q}\}} && \min_{q} \quad {\mathrm{SINR}}_q && \\
    & \text{subject to}     && \mathrm{SINR}_u \geq \gamma_u, \quad \forall u,\label{eq:st_SINR_u}\\
    &  && \frac{1}{L}\sum_{l} \|\mathbf{x}[l]\|^2 \leq P_{\mathrm{max}},\label{eq:st_power}\\ 
    &  && \mathbf{w}_{q}^{\H}\tilde{\mathbf{A}}_{q}\mathbf{x} = 1, && \label{eq:st_Rx_filter}
\end{align}
\end{subequations}
where \eqref{eq:st_SINR_u} ensures that each communication user achieves a predetermined SINR threshold $\gamma_u$, and \eqref{eq:st_power} enforces the transmit power constraint, with $P_{\mathrm{max}}$ denoting the maximum power per time slot. Similarly, when the transmitted symbols are unknown to the sensing receiver, we have
\begin{subequations}\label{eq:ProblemFormulationMain_avg}
\begin{align}
    (\mathbf{P2}) ~~~&  \max_{\mathbf{F},\{\mathbf{w}_{q}\}} && \min_{q} \quad {\overline{\mathrm{SINR}}}_q && \\
    & \text{subject to}     && \overline{\mathrm{SINR}}_u \geq \gamma_u, \quad \forall u,\label{eq:st_SINR_u_avg}\\
    &  && \sum_{u} \|\mathbf{f}_u\|^2 \leq P_{\mathrm{max}},\label{eq:st_power_avg}\\ &
    && \|\mathbf{w}_{q}\|^2 = 1. && \label{eq:st_Rx_filter_avg}
\end{align}
\end{subequations}

\section{Joint Design of Beamforming and Rx Filters}\label{sec:proposedalg}
In this section, we propose  solutions for the cases where the transmit (Tx) symbols are known (problem P1) or unknown (problem P2) to the sensing receiver.

\subsection{Known Transmit Symbols: Solving Problem P1}\label{sec:proposed_known}
% We first derive the optimal receive  filters in closed form, then formulate a tractable beamforming design using successive convex approximation (SCA) and fractional programming (FP). Specifically, for a given 
% beamforming matrix
% $\mathbf{F}$, the optimal Rx filter $\mathbf{w}_q$ is obtained via the minimum variance distortionless response (MVDR) problem:
We first derive the optimal receive filters in closed form and then develop a tractable beamforming design using successive convex approximation (SCA) and fractional programming (FP). For a given $\mathbf{F}$, the optimal filter is obtained via the minimum variance distortionless response (MVDR) formulation:
\begin{align}\label{eq:ProblemFormulation_RxFilter}
    \min_{\mathbf{w}_q} \quad \mathbf{w}_{q}^{\H}\mathbf{R}_{q}\mathbf{w}_{q}, ~~ \text{subject to}~ \eqref{eq:st_Rx_filter}.
\end{align}
A solution to \eqref{eq:ProblemFormulation_RxFilter} is given by \cite[(15)]{Satyabrata2014_ofdmRadarSTAP} as
\begin{equation}\label{eq:opt_rx_filter}
    \mathbf{w}_q^\star = \alpha \mathbf{R}_{q}^{-1}\tilde{\mathbf{A}}_q \mathbf{x},
\end{equation}
where $\alpha = \big( \mathbf{x}^{\H}\tilde{\mathbf{A}}_q^{\H} \mathbf{R}_{q}^{-1} \tilde{\mathbf{A}}_q\mathbf{x} \big)^{-1}$ is a normalization factor ensuring \eqref{eq:st_Rx_filter}.
Substituting \eqref{eq:opt_rx_filter} into \eqref{eq:SINR_radar} yields 
\begin{equation}\label{eq:SINR_radar_replacing_w}
    \mathrm{SINR}^{\Rad}_q = |\beta_{q}|^2 \mathbf{x}^{\H}\tilde{\mathbf{A}}_q^{\H} \mathbf{R}_{q}^{-1} \tilde{\mathbf{A}}_q\mathbf{x}.
\end{equation}
Therefore, \eqref{eq:ProblemFormulationMain} reduces to
\begin{equation}\label{eq:ProblemFormulationMainReduced}
    \max_{\mathbf{F}}\quad \min_{q} \mathrm{SINR}^{\Rad}_q, \quad \text{subject to}  ~\eqref{eq:st_SINR_u} - \eqref{eq:st_power}.
\end{equation}
Introducing 
% a slack variable 
$\tau \in \mathbb{R}$, \eqref{eq:ProblemFormulationMainReduced} can be written in epigraph form as
\begin{subequations}\label{eq:ProblemFormulationEpigraph}
\begin{align}
    &  \max_{\mathbf{F},\tau} && \tau && \\
    & \text{subject to} && {\mathrm{SINR}}_q \geq \tau,~\forall q,\label{eq:st_SINR_q}\\
    &  && \eqref{eq:st_SINR_u} - \eqref{eq:st_power}.
\end{align}
\end{subequations}
The problem remains non-convex due to \eqref{eq:st_SINR_q} and \eqref{eq:st_SINR_u}. To handle \eqref{eq:st_SINR_u}, we introduce $\xi_u[l] > 0$ and rewrite \eqref{eq:st_SINR_u} as
\begin{subequations}\label{eq:rewrite_st_SINR_u}
\begin{align}
    & \frac{1}{L}\sum_{l} \xi_{u}[l] \geq \gamma_u,~\forall u,\label{eq:st_sum_xi}\\
    & \sum_{j = 1, j\neq u}^{U} |\tilde{\mathbf{h}}_{u}^{\H} \mathbf{f}_{j}s_{j}[l]|^2 + 1
    + f_{\text{qol}}(\mathbf{f}_{u},\xi_{u}[l]) 
    \leq 0.\label{eq:st_SINRuk_xi}
\end{align}
\end{subequations}
where we define $f_{\text{qol}}(\mathbf{f}_{u},\xi_{u}[l]) \triangleq -\frac{|\tilde{\mathbf{h}}_{u}^{\H} \mathbf{f}_{u}s_{u}[l]|^2}{\xi_{u}[l]}$. We apply SCA, where in each iteration $i$, $f_{\text{qol}}$ is replaced by a tight convex upper bound using a first-order Taylor expansion, i.e., 
\begin{align}
        & f_{\text{qol}}(\mathbf{f}_{u},\xi_{u}[l]) \leq F_{\text{qol}}(\mathbf{f}_{u},\xi_{u}[l],\mathbf{f}_{u}^{(i)},\xi_{u}^{(i)}[l]) = \notag\\
        & \frac{(\mathbf{f}_{u}^{(i)})^{\H} \tilde{\mathbf{h}}_{u}\tilde{\mathbf{h}}_{u}^{\H}\mathbf{f}_{u}^{(i)}}{(\xi_{u}^{(i)}[l])^{2}} \xi_{u}[l]
        - \frac{2\Re\left\{(\mathbf{f}_{u}^{(i)})^{\H} \tilde{\mathbf{h}}_{u}\tilde{\mathbf{h}}_{u}^{\H}\mathbf{f}_{u}\right\}}{\xi_{u}^{(i)}[l]}.
\end{align}
Accordingly, \eqref{eq:st_SINRuk_xi} is approximated as \begin{align}\label{eq:st_SINRuk_xi_convex}
    \sum_{j\neq u} |\tilde{\mathbf{h}}_{u}^{\H} \mathbf{f}_{j}s_{j}[l]|^2 + 1 +F_{\text{qol}}(\mathbf{f}_{u},\xi_{u}[l],\mathbf{f}_{u}^{(i)},\xi_{u}^{(i)}[l]) \leq 0.
\end{align}

% \red{Next, to deal with the non-convexity of \eqref{eq:st_SINR_q}, we can apply the quadratic transformation \cite[Sec.~D]{shen2018fractinal} and rewrite \eqref{eq:st_SINR_q} as}
To handle \eqref{eq:st_SINR_q}, we apply the quadratic transform \cite{shen2018fractinal}. Introducing
$\mathbf{y}_{q} \triangleq \mathbf{R}_{q}^{-1} \tilde{\mathbf{A}}_{q} \tilde{\mathbf{F}} \mathbf{s}$, we rewrite \eqref{eq:st_SINR_q} as
\begin{equation}
    2\Real\{\mathbf{y}_{q}^\H \tilde{\mathbf{A}}_{q} \tilde{\mathbf{F}} \mathbf{s}\} -  \mathbf{y}_{q}^\H \mathbf{R}_{q}\mathbf{y}_{q}\geq \tau,~\forall q.\label{eq:st_SINR_q_FP}
\end{equation}
% where $\mathbf{y}_{q} \triangleq \mathbf{R}_{q}^{-1} \tilde{\mathbf{A}}_{q} \tilde{\mathbf{F}} \mathbf{s}$. 
The term $\mathbf{y}_{q}^{\H}\mathbf{R}_{q}\mathbf{y}_{q}$ can be expressed in the quadratic form as
\begin{equation}
    \mathbf{y}_{q}^{\H}\mathbf{R}_{q}\mathbf{y}_{q} = \tilde{\mathbf{f}}^{\H} \mathbf{G}_{q}\tilde{\mathbf{f}} + \sigma_R^2 \mathbf{y}_{q}^\H \mathbf{y}_{q},
\end{equation}
where $\mathbf{G}_{q}\!\triangleq \!\!\sum_{\bar q \neq q} |\beta_{\bar q}|^2 \!((\mathbf{s}\mathbf{s}^\H)^\tran \!\otimes\tilde{\mathbf{A}}_{\bar{q}}^\H\mathbf{y}_{q}\mathbf{y}_{q}^\H \tilde{\mathbf{A}}_{\bar{q}}\!)$ and
$\tilde{\mathbf{f}} \!\triangleq\! \mathrm{vec}(\tilde{\mathbf{F}})$.
Hence, \eqref{eq:st_SINR_q} becomes the convex quadratic inequality 
\begin{equation}\label{eq:st_SINR_q_FP_convex}
    2\Real\{\mathbf{y}_{q}^\H \tilde{\mathbf{A}}_{q} \tilde{\mathbf{F}} \mathbf{s}\} 
    - \tilde{\mathbf{f}}^\H \mathbf{G}_{q} \tilde{\mathbf{f}} 
    - \sigma_R^2 \mathbf{y}_{q}^\H \mathbf{y}_{q} \geq \tau.
\end{equation}

Finally, the resulting convex problem at iteration $i$ is
\begin{equation}\label{eq:ProblemFormulationBeamforming_convexified}
     \max_{\mathbf{F},\tau,\{\xi_u[l]\}}\quad \tau, \quad \text{subject to} \eqref{eq:st_power}, \eqref{eq:st_sum_xi},\eqref{eq:st_SINRuk_xi_convex}, \eqref{eq:st_SINR_q_FP_convex}. 
\end{equation}
% \begin{subequations}\label{eq:ProblemFormulationBeamforming_convexified}
% \begin{align}
%     &  \max_{\mathbf{F},\tau} && \tau  && \\
%     & \text{subject to}     &&  \eqref{eq:st_power}, \eqref{eq:st_sum_xi},\eqref{eq:st_positive_xi},\eqref{eq:st_SINRuk_xi_convex}, \eqref{eq:st_SINR_q_FP_convex}.
% \end{align}
% \end{subequations}
To avoid infeasibility in early iterations, we introduce an auxiliary variable $\eta$ and redefine constraint \eqref{eq:st_sum_xi} as
\begin{equation}
    \frac{1}{L}\sum_{l} \xi_{u}[l] \geq \gamma_u - \eta,~\forall u,\label{eq:st_sum_xi_eta}
\end{equation}
allowing \eqref{eq:ProblemFormulationBeamforming_convexified} to be rewritten as
\begin{equation}\label{eq:ProblemFormulationBeamforming_convexified_eta}
     \max_{\mathbf{F},\tau,\{\xi_u[l]\},\eta}~ \tau  + \alpha_\eta^{(i)}|\eta|, ~\text{subject to}~ \eqref{eq:st_power}, \eqref{eq:st_sum_xi_eta},\eqref{eq:st_SINRuk_xi_convex}, \eqref{eq:st_SINR_q_FP_convex},
\end{equation}
where $\alpha_\eta^{(i)}$ is a penalty factor that discourages violations of \eqref{eq:st_sum_xi}. At each iteration, $\alpha_\eta^{(i)}$ is increased to drive $\eta$ toward zero at convergence. If $\eta$ remains large after convergence, the problem is infeasible for the given $\gamma_q$, and $\gamma_q - \eta$ represents the best achievable communications SINR.

Finally, the beamforming design is obtained using Algorithm~\ref{alg:Beamforming}. At each iteration $i$, the convex problem \eqref{eq:ProblemFormulationBeamforming_convexified_eta} is solved with standard convex optimization tools such as CVX.

\begin{algorithm}[t]
\caption{Joint beamforming and Rx filter design via SCA-FP for known Tx symbols (problem P1)}
\label{alg:Beamforming}
\begin{algorithmic}
\STATE \textbf{Initialize:} $\mathbf{F}^{(0)}$, $\xi_{u}^{(0)}[l]$, $i = 0$, and set a small threshold $\epsilon$.
\REPEAT
\STATE Solve \eqref{eq:ProblemFormulationBeamforming_convexified_eta} to obtain $\tau^{\star}$, $\mathbf{F}^{\star}$, $\{\xi_{u}^{\star}[l]\}$, and $\eta^\star$.
\STATE Update $\mathbf{F}^{(i+1)} = \mathbf{F}^{\star}$, $\tau^{(i+1)} = \tau^{\star}$, $\xi_{u}^{(i+1)} = \xi_{u}^{\star}$.
\STATE Set $i \leftarrow i+1$.
\UNTIL{$|\tau^{(i+1)} - \tau^{(i)}| \leq \epsilon$ \textbf{and} $|\eta^\star| \approx 0$}.
\STATE \textbf{Output:} Beamforming $\mathbf{F}$ and Filters $\mathbf{w}_q = \alpha \mathbf{R}_{q}^{-1}\tilde{\mathbf{A}}_q \tilde{\mathbf{F}}\mathbf{s}$.
\end{algorithmic}
\end{algorithm}

% % [MODIFY tehbelwo algoeithm as well]
% \begin{algorithm}[t]
% % \begin{small}
% \caption{Solution to the beamforming design in \eqref{eq:ProblemFormulationWaveform}}
% \label{alg:Beamforming}
% \begin{algorithmic}
% \STATE \textbf{initialize}:  $\mathbf{F}_k^{(0)}$ and $\xi_{u,k}^{(0)}$, $i = 0$, and a small threshold $\epsilon_1$.
% \REPEAT
% \STATE Obtain $\{\tau^{\star},\{\mathbf{F}_k^{\star}\},\{\xi_{u,k}^{\star}\},\eta^\star\}$ by solving \eqref{eq:ProblemFormulationBeamforming_convexified_eta} (or simplified version \eqref{eq:ProblemFormulationBeamforming_convexified_eta_simplified}). 
% \STATE Update $\mathbf{F}_k^{(i+1)} = \mathbf{F}_k^{\star}$, $\tau^{(i+1)} = \tau^{\star}$, $\xi_{u,k}^{(i+1)} = \xi_{u,k}^{\star}$ and set $i = i+1$.
% \UNTIL{$|\tau^{(i+1)} - \tau^{(i)}| \leq \epsilon_1 $ and $|\eta^\star| \approx 0$}.
% \STATE \textbf{output}: Beamforming matrices $\mathbf{F}_k$, $\forall k$.
% \end{algorithmic}
% % \end{small}
% \end{algorithm}

\subsection{Unknown Transmit Symbols: Solving Problem P2}\label{sec:proposed_unknown}
When the transmit symbols are unknown at the radar receiver, directly substituting the Rx filters into the sensing SINR, as in \eqref{eq:SINR_radar_replacing_w}, is not straightforward. We therefore employ alternating optimization (AO), iteratively solving the beamforming and filter subproblems while keeping the other fixed.

\subsubsection{Beamforming Design} Given $\{\mathbf{w}_q\},~\forall q$, the beamforming subproblem is rewritten in the following epigraph form:
\begin{subequations}\label{eq:ProblemFormulationEpigraphAvg}
\begin{align}
    &  \max_{\mathbf{F},\tau} && \tau && \\
    & \text{subject to} && \overline{\mathrm{SINR}}_q \geq \tau,~\forall q,\label{eq:st_SINR_q_avg}\\
    &  && \eqref{eq:st_SINR_u_avg} - \eqref{eq:st_power_avg}.
\end{align}
\end{subequations}
Constraints  \eqref{eq:st_SINR_q_avg} and \eqref{eq:st_SINR_u_avg} are non-convex. We adopt a similar approach as in Section~\ref{sec:proposed_known} and rewrite \eqref{eq:st_SINR_u_avg} as
\begin{subequations}\label{eq:rewrite_st_SINR_u_avg}
\begin{align}
    & \xi_{u} \geq \gamma_u - \eta,~\forall u,\label{eq:st_sum_xi_avg}\\
    &     \sum_{j\neq u} |\tilde{\mathbf{h}}_{u}^{\H} \mathbf{f}_{j}|^2 + 1 +F_{\text{qol}}(\mathbf{f}_{u},\xi_{u},\mathbf{f}_{u}^{(i)},\xi_{u}^{(i)}) \leq 0.\label{eq:st_SINRuk_xi_avg}
\end{align}
\end{subequations}

To handle the non-convexity of \eqref{eq:st_SINR_q_avg}, it is rewritten as
\begin{align}
    |\beta_{q}|^2 f_q(\mathbf{F}) - 
    \sum_{\bar{q} \neq q} |\beta_{\bar{q}}|^2 g_{q,\bar{q}}(\mathbf{F},\tau)
    + \tau\sigma_R^2 \mathbf{w}_q^{\mathrm{H}}\mathbf{w}_q \geq 0,
\end{align}
where
\begin{align}
    f_q(\mathbf{F}) &\triangleq \mathbf{w}_{q}^{\H}\tilde{\mathbf{A}}_{q}\tilde{\mathbf{F}}\tilde{\mathbf{F}}^{\H}\tilde{\mathbf{A}}_{q}^{\H}\mathbf{w}_{q}, \\
    g_{q,\bar{q}}(\mathbf{F},\tau) &\triangleq \tau \mathbf{w}_q^{\H} \tilde{\mathbf{A}}_{\bar{q}}\tilde{\mathbf{F}}\tilde{\mathbf{F}}^{\H} \tilde{\mathbf{A}}_{\bar{q}}^{\H} \mathbf{w}_q.
\end{align}
% where $f_q(\mathbf{F}) \triangleq \mathbf{w}_{q}^{\mathrm{H}}\tilde{\mathbf{A}}_{q}\tilde{\mathbf{F}}\tilde{\mathbf{F}}^{\mathrm{H}}\tilde{\mathbf{A}}_{q}^{\mathrm{H}}\mathbf{w}_{q}$  and $g_{q,\bar{q}}(\mathbf{F},\tau) \triangleq  \tau \mathbf{w}_q^{\mathrm{H}} \tilde{\mathbf{A}}_{\bar{q}}\tilde{\mathbf{F}}\tilde{\mathbf{F}}^{\mathrm{H}} \tilde{\mathbf{A}}_{\bar{q}}^{\mathrm{H}} \mathbf{w}_q$.
We apply SCA and approximate $f_q$ and $g_{q,\bar{q}}$ via first-order Taylor expansions around $(\mathbf{F}^{(i)})$ and $(\mathbf{F}^{(i)},\tau^{(i)})$, respectively:
\begin{align}
    & \hat{f}_q(\mathbf{F}) 
    = f_q(\mathbf{F}^{(i)}) 
    + 2\Real\left\{\mathbf{w}_q^{\H}\tilde{\mathbf{A}}_q \tilde{\mathbf{F}}^{(i)} (\tilde{\mathbf{F}} - \tilde{\mathbf{F}}^{(i)})^{\H} \tilde{\mathbf{A}}_q^{\H}\mathbf{w}_q \right\},\notag \\
    & \hat{g}_{q,\bar{q}}(\mathbf{F},\tau) 
    \!=\! g_{q,\bar{q}}(\mathbf{F}^{(i)},\tau^{(i)}\!) \!+\! (\tau \!- \!\tau^{(i)}) \mathbf{w}_q^{\H}\tilde{\mathbf{A}}_{\bar{q}}\tilde{\mathbf{F}}^{(i)}\tilde{\mathbf{F}}^{(i)\H}\tilde{\mathbf{A}}_{\bar{q}}^{\H}\mathbf{w}_q \notag \\
    &\quad\quad + 2\tau^{(i)} \Real\left\{\mathbf{w}_q^{\H}\tilde{\mathbf{A}}_{\bar{q}}\tilde{\mathbf{F}}^{(i)}(\tilde{\mathbf{F}} - \tilde{\mathbf{F}}^{(i)})^{\H}\tilde{\mathbf{A}}_{\bar{q}}^{\H}\mathbf{w}_q \right\}\notag.
\end{align}
Accordingly, the constraint is approximated by the following convex linear form
\begin{align}\label{eq:st_SINR_q_avg_convex}
    |\beta_q|^2 \hat{f}_q(\mathbf{F}) 
    - \sum_{\bar{q} \neq q} |\beta_{\bar{q}}|^2 \hat{g}_{q,\bar{q}}(\mathbf{F},\tau)
    + \tau\sigma_R^2 \mathbf{w}_q^{\mathrm{H}}\mathbf{w}_q \geq 0.
\end{align}
\begin{algorithm}[t!]
\caption{AO-based joint beamforming and Rx filter design for unknown Tx symbols (problem P2)}
\label{alg:AO}
\begin{algorithmic}
\STATE \textbf{Initialize:} $\mathbf{F}^{(0)}$, $\{\mathbf{w}_q^{(0)}\}$, $\{\xi_{u}^{(0)}\}$, tolerance $\epsilon$, and $i=0$.
\REPEAT
\STATE  Update $\tau^{(i+1)}$ and $\mathbf{F}^{(i+1)}$ by solving \eqref{eq:ProblemFormulationBeamforming_convexified_eta_avg}.
\STATE Update $\mathbf{w}_q^{(i+1)}$, $\forall q$, using \eqref{eq:update_wq_avg} and $i \leftarrow i+1$.
\UNTIL{$|\tau^{(i+1)} - \tau^{(i)}| \leq \epsilon$.}
\STATE \textbf{Output:} Beamforming $\mathbf{F}$ and Filters $\mathbf{w}_q$.
\end{algorithmic}
\end{algorithm}
Finally, the convexified beamforming subproblem is given by
\begin{equation}\label{eq:ProblemFormulationBeamforming_convexified_eta_avg}
     \hspace{-1mm}\max_{\mathbf{F},\tau,\{\xi_u\},\eta}~ \tau  + \alpha_\eta^{(i)}|\eta|, ~\text{subject to}~ \eqref{eq:st_power}, \eqref{eq:st_sum_xi_avg},\eqref{eq:st_SINRuk_xi_avg}, \eqref{eq:st_SINR_q_avg_convex}.
\end{equation}

\subsubsection{Rx Filter Design}
For a given beamforming matrix $\mathbf{F}$, the Rx filter subproblem for each $q$ is given by
\begin{equation}\label{eq:ProblemFormulationFilterDesignAvg}
    \max_{\mathbf{w}_q}\quad 
    \frac{\mathbf{w}_q^{\H}\tilde{\mathbf{A}}_{q}\tilde{\mathbf{F}}\tilde{\mathbf{F}}^{\H}\tilde{\mathbf{A}}_{q}^{\H}\mathbf{w}_q}
    {\mathbf{w}_q^{\H}\bar{\mathbf{R}}_q\mathbf{w}_q},\quad \text{subject to}  ~\eqref{eq:st_Rx_filter_avg}.
\end{equation}
Problem \eqref{eq:ProblemFormulationFilterDesignAvg} is a generalized Rayleigh quotient and its optimal solution is given by the dominant generalized eigenvector, i.e.,
\begin{equation}\label{eq:update_wq_avg}
    \mathbf{w}_q^{\star} = \alpha\, \bar{\mathbf{R}}_q^{-1}\mathbf{u}_{\max},
\end{equation}
where $\mathbf{u}_{\max}$ is the principal eigenvector of 
$\bar{\mathbf{R}}_q^{-1}\tilde{\mathbf{A}}_q\tilde{\mathbf{F}}\tilde{\mathbf{F}}^{\H}\tilde{\mathbf{A}}_q^{\H}$. The corresponding maximum SINR is
\begin{equation}
    \mathrm{SINR}^{\Rad}_q  
    = \lambda_{\max}\left(
    \bar{\mathbf{R}}_q^{-1}
    \tilde{\mathbf{A}}_q
    \tilde{\mathbf{F}}\tilde{\mathbf{F}}^{\H}
    \tilde{\mathbf{A}}_q^{\H}
    \right).
\end{equation}

Based on the above, the overall AO procedure for solving P2 is summarized in Algorithm~\ref{alg:AO}.

\section{Numerical Results}\label{sec:results}

\begin{figure}[t]
\newcommand\figwidth{0.7}
\centering
\resizebox{\figwidth\columnwidth}{!}{
\definecolor{mycolor1}{rgb}{1.00000,0.00000,1.00000}
\begin{tikzpicture}

\begin{axis}[%
width=4.521in,
height=3.566in,
at={(0.758in,0.481in)},
scale only axis,
xmin=0,
xmax=100,
xlabel style={labelLargefont},
xlabel={X-axis},
ymin=-24.8673766630053,
ymax=54.0035910789301,
ylabel style={labelLargefont},
ylabel={Y-axis},
axis background/.style={fill=white},
title style={font=\bfseries},
% axis x line*=bottom,
% axis y line*=left,
xmajorgrids,
ymajorgrids,
legend style={legend cell align=left, align=left, draw=white!15!black}
]
\addplot [color=blue, line width=2.0pt, only marks, mark size=4.0pt, mark=o, mark options={solid, blue}] table[]{\datafolder Fig_Bistatic_Scenario-1.tsv};
\addlegendentry{Tx}
\node[right, align=left, inner sep=0] at (axis cs:0,-3) { Tx};

\addplot [color=red, line width=2.0pt, only marks, mark size=4.0pt, mark=o, mark options={solid, red}]table[]{\datafolder Fig_Bistatic_Scenario-2.tsv};
\addlegendentry{Rx}
\node[right, align=left, inner sep=0] at (axis cs:96,-3) {Rx};

\addplot [color=black, line width=2.0pt] table[]{\datafolder Fig_Bistatic_Scenario-3.tsv};
\addlegendentry{LOS}
\node[right, align=left, inner sep=0] at (axis cs:50,2) { LOS};
\addplot [color=mycolor1, only marks, mark size=6.0pt, mark=asterisk, mark options={solid, mycolor1}] table[]{\datafolder Fig_Bistatic_Scenario-4.tsv};
\addlegendentry{Targets}

\addplot [color=green, line width=1.5pt, only marks, mark size=2.8pt, mark=square, mark options={solid, green}] table[]{\datafolder Fig_Bistatic_Scenario-5.tsv};
\addlegendentry{Users}

\node[right, align=left, inner sep=0] at (axis cs:49.244,-12.268) {$\text{U}_\text{1}$};
\node[right, align=left, inner sep=0] at (axis cs:11.932,49.657) {$\text{U}_\text{2}$};
\node[right, align=left, inner sep=0] at (axis cs:15.766,-9.18) {$\text{U}_\text{3}$};
\node[right, align=left, inner sep=0] at (axis cs:35.486,27.189) {$\text{U}_\text{4}$};

\addplot [color=blue, dashed, line width=1.2pt] table[]{\datafolder Fig_Bistatic_Scenario-6.tsv};
\addlegendentry{Tx paths}

\addplot [color=blue, dashed, line width=1.2pt] table[]{\datafolder Fig_Bistatic_Scenario-7.tsv};
\addlegendentry{Rx paths}

\node[right, align=left, inner sep=0]
at (axis cs:56.382,-20.521) {$\text{T}_\text{1}$};

% \addplot [color=blue, forget plot] table[]{\datafolder Fig_Bistatic_Scenario-8.tsv};
% \node[right, align=left, inner sep=0] at (axis cs:14.772,-2.605) {$\phi{}_{\text{1}}\text{=-20.0}^\circ$};

\addplot [color=blue, dashed, line width=1.2pt, forget plot] table[]{\datafolder Fig_Bistatic_Scenario-9.tsv};
\addplot [color=blue, dashed, line width=1.2pt, forget plot] table[]{\datafolder Fig_Bistatic_Scenario-10.tsv};

\node[right, align=left, inner sep=0] at (axis cs:68.937,12.155) {$\text{T}_\text{2}$};

% \addplot [color=blue, forget plot] table[]{\datafolder Fig_Bistatic_Scenario-11.tsv};
% \node[right, align=left, inner sep=0] at (axis cs:14.943,1.307) {$\phi{}_{\text{2}}\text{=10.0}^\circ$};

\addplot [color=blue, dashed, line width=1.2pt, forget plot]
  table[]{\datafolder Fig_Bistatic_Scenario-12.tsv};
\addplot [color=blue, dashed, line width=1.2pt, forget plot]
  table[]{\datafolder Fig_Bistatic_Scenario-13.tsv};
\node[right, align=left, inner sep=0]
at (axis cs:58.91,27.47) {$\text{T}_\text{3}$};

% \addplot [color=blue, forget plot] table[]{\datafolder Fig_Bistatic_Scenario-14.tsv};
% \node[right, align=left, inner sep=0] at (axis cs:14.772,-2.605) {$\phi{}_{\text{1}}\text{=-20.0}^\circ$};

% \node[right, align=left, inner sep=0] at (axis cs:14.943,1.307) {$\phi{}_{\text{2}}\text{=10.0}^\circ$};

% \node[right, align=left, inner sep=0] at (axis cs:14.644,3.247) {$\phi{}_{\text{3}}\text{=25.0}^\circ$};

% \draw[thick]
% (axis cs:10,0) arc[start angle=0, end angle=10, radius=8];
% \node at (axis cs:6,2) {$\phi_1$};

% \draw[thick]
% (axis cs:8,0) arc[start angle=0, end angle=-20, radius=8];
% \node at (axis cs:15,4) {$\phi_2$};

% % Draw angle arc near Tx
% \draw[thick]
% (axis cs:15,0) arc[start angle=0, end angle=25, radius=8];
% \node at (axis cs:15,4) {$\phi_3$};

\end{axis}
\end{tikzpicture}
}
\vspace{-3mm}
\caption{Example of the considered simulation setup.}
\label{fig:Bistatic_Scenario}
\end{figure}
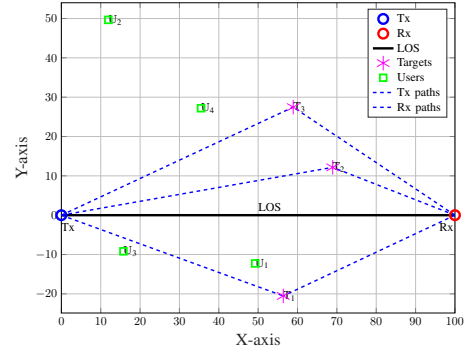
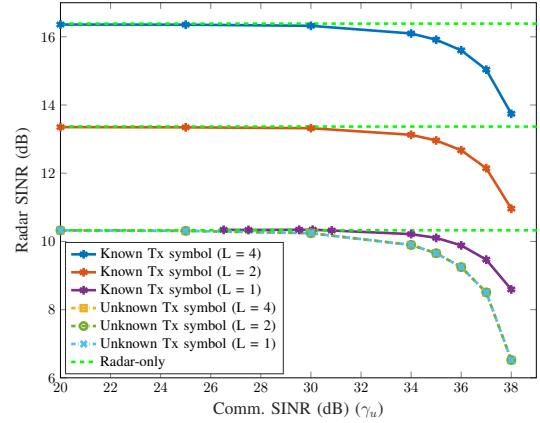
\begin{figure}[t]
\newcommand\figwidth{0.8}
\centering
\resizebox{\figwidth\columnwidth}{!}{
\definecolor{mycolor1}{rgb}{0.00000,0.44700,0.74100}%
\definecolor{mycolor2}{rgb}{0.85000,0.32500,0.09800}%
\definecolor{mycolor3}{rgb}{0.92900,0.69400,0.12500}%
\definecolor{mycolor4}{rgb}{0.49400,0.18400,0.55600}%
\definecolor{mycolor5}{rgb}{0.46600,0.67400,0.18800}%
\definecolor{mycolor6}{rgb}{0.30100,0.74500,0.93300}%
\definecolor{mycolor7}{rgb}{0.63500,0.07800,0.18400}%
\begin{tikzpicture}

\begin{axis}[%
width=4.521in,
height=3.566in,
at={(0.758in,0.481in)},
scale only axis,
xmin=20,
xmax=39,
xlabel style={labellargefont},
xlabel={Comm. SINR (dB) ($\gamma_u$)},
ymin=6,
ymax=17,
ylabel style={labellargefont},
ylabel={Radar SINR (dB)},
axis background/.style={fill=white},
legend style={at={(0.01,0.01)}, anchor=south west, legend cell align=left, align=left, draw=white!15!black}
]

\addplot [color=mycolor1, line width=1.8pt, mark=asterisk, mark size=3pt, mark options={solid, mycolor1}]
  table[]{\datafolder Fig_L_Bistatic-4.tsv};
\addlegendentry{Known Tx symbol (L = 4)}

\addplot [color=mycolor2, line width=1.8pt, mark=asterisk, mark size=3pt, mark options={solid, mycolor2}]
  table[]{\datafolder Fig_L_Bistatic-5.tsv};
\addlegendentry{Known Tx symbol (L = 2)}

\addplot [color=mycolor4, line width=1.8pt, mark=asterisk, mark size=3pt, mark options={solid, mycolor4}]
  table[]{\datafolder Fig_L_Bistatic-6.tsv};
\addlegendentry{Known Tx symbol (L = 1)}

\addplot [color=mycolor3, dashdotted, line width=1.8pt, mark=square, mark options={solid, mycolor3}]
  table[]{\datafolder Fig_L_Bistatic-7.tsv};
\addlegendentry{Unknown Tx symbol (L = 4)}

\addplot [color=mycolor5, dashdotted, line width=1.8pt, mark=o, mark size=2.5pt, mark options={solid, mycolor5}]
  table[]{\datafolder Fig_L_Bistatic-8.tsv};
\addlegendentry{Unknown Tx symbol (L = 2)}

\addplot [color=mycolor6, dashdotted, line width=1.8pt, mark=x, mark size=3pt,  mark options={solid, mycolor6}]
  table[]{\datafolder Fig_L_Bistatic-9.tsv};
\addlegendentry{Unknown Tx symbol (L = 1)}

\addplot [color=green, dashed, line width=1.8pt]
  table[]{\datafolder Fig_L_Bistatic-1.tsv};
\addlegendentry{Radar-only}

\addplot [color=green, dashed, line width=1.8pt]
  table[]{\datafolder Fig_L_Bistatic-2.tsv};

\addplot [color=green, dashed, line width=1.8pt]
  table[]{\datafolder Fig_L_Bistatic-3.tsv};

\end{axis}
\end{tikzpicture}
}
\vspace{-3mm}
\caption{Radar-communications performances for different frame length $L$.}
\label{fig:Bistatic_L}
\end{figure}
\begin{figure}[t]
\newcommand\figwidth{0.75}
\centering
\resizebox{\figwidth\columnwidth}{!}{
\definecolor{mycolor1}{rgb}{0.00000,0.44700,0.74100}%
\definecolor{mycolor2}{rgb}{0.85000,0.32500,0.09800}%
\definecolor{mycolor3}{rgb}{0.92900,0.69400,0.12500}%
\definecolor{mycolor4}{rgb}{0.49400,0.18400,0.55600}%
\definecolor{mycolor5}{rgb}{0.46600,0.67400,0.18800}%
\definecolor{mycolor6}{rgb}{0.30100,0.74500,0.93300}%
\definecolor{mycolor7}{rgb}{0.63500,0.07800,0.18400}%
\begin{tikzpicture}

\begin{axis}[%
width=4.521in,
height=3.566in,
at={(0.758in,0.481in)},
scale only axis,
xmin=20,
xmax=39,
xlabel style={labellargefont},
xlabel={Comm. SINR (dB) ($\gamma_u$)},
ymin=0,
ymax=18,
ylabel style={labellargefont},
ylabel={Radar SINR (dB)},
axis background/.style={fill=white},
legend style={at={(0.01,0.01)}, anchor=south west, legend cell align=left, align=left, draw=white!15!black}
]
\addplot [color=green, dashed, line width=1.8pt]
  table[]{\datafolder Fig_U_Bistatic-1.tsv};
\addlegendentry{Radar-only (known Tx symbol)}

\addplot [color=mycolor1, line width=1.8pt, mark=asterisk, mark size=2.5pt, mark options={solid, mycolor1}]
  table[]{\datafolder Fig_U_Bistatic-2.tsv};
\addlegendentry{Known Tx symbolmod (U = 2)}

\addplot [color=mycolor2, line width=1.8pt, mark=asterisk, mark size=2.5pt, mark options={solid, mycolor2}]
  table[]{\datafolder Fig_U_Bistatic-3.tsv};
\addlegendentry{Known Tx symbolmod (U = 3)}

\addplot [color=mycolor3, line width=1.8pt, mark=asterisk, mark size=2.5pt, mark options={solid, mycolor3}]
  table[]{\datafolder Fig_U_Bistatic-4.tsv};
\addlegendentry{Known Tx symbolmod (U = 5)}

\addplot [color=mycolor4, dashed, line width=1.8pt]
  table[]{\datafolder Fig_U_Bistatic-5.tsv};
\addlegendentry{Radar-only (unknown Tx symbol)}

\addplot [color=mycolor5, dashdotted, line width=1.8pt, mark=square, mark options={solid, mycolor5}]
  table[]{\datafolder Fig_U_Bistatic-6.tsv};
\addlegendentry{Unknown Tx symbol (U = 2)}

\addplot [color=mycolor6, dashdotted, line width=1.8pt, mark=square, mark options={solid, mycolor6}]
  table[]{\datafolder Fig_U_Bistatic-7.tsv};
\addlegendentry{Unknown Tx symbol (U = 3)}

\addplot [color=mycolor1, dashdotted, line width=1.8pt, mark=square, mark options={solid, mycolor1}]
  table[]{\datafolder Fig_U_Bistatic-8.tsv};
\addlegendentry{Unknown Tx symbol (U = 5)}

\end{axis}
\end{tikzpicture}
}
\vspace{-3mm}
\caption{Radar-communications performances for different number of communications users $U$, with $L = 4$.}
\label{fig:Bistatic_U}
\end{figure}
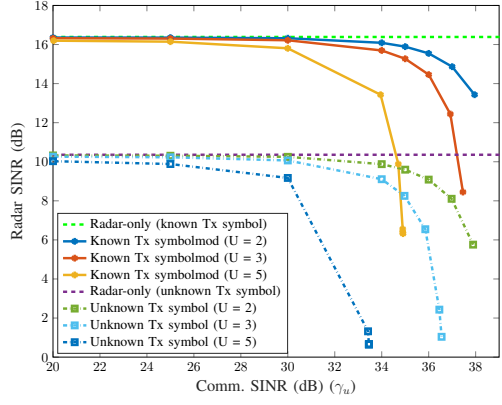
We present simulation results to evaluate the proposed algorithms. The simulation parameters are as follows:
$P_{\mathrm{max}} = 1~\mathrm{W}$, $N = M = 16$, $f_c = 1~\mathrm{GHz}$, $\sigma^2_{\mathrm{C}} = \sigma^2_{\mathrm{R}} = 0.001$, and the antenna spacing at both the transmit and receive arrays is $d = \lambda_c/2$. We consider $Q = 3$ targets located at angles $\phi_q \in \{10^\circ, -20^\circ, 25^\circ\}$ and $\theta_q \in \{158.6^\circ, -154.8^\circ, 146.2^\circ\}$, with corresponding range bins $l_q \in \{52, 54, 57\}$ for $q = 1, \ldots, Q$, and reflection coefficients $|\beta_q| = -15~\mathrm{dB}$, $\forall q$. The channel gain of the LOS is set to $\beta_0 = -5~\mathrm{dB}$. The convergence tolerance is chosen as $\epsilon = 0.005$.
Further, the communication channels are assumed to follow Rayleigh fading model. For the benchmarking, we consider a radar-only design obtained by solving \eqref{eq:ProblemFormulationMain} (or \eqref{eq:ProblemFormulationMain_avg}) without imposing the communication SINR constraints. The minimum sensing SINR achieved by the radar-only design thus serves as an upper bound on the sensing performance of the proposed ISAC schemes.

Fig.~\ref{fig:Bistatic_L} illustrates the radar--communications tradeoff for different frame lengths $L$ when the transmitted signals are either known or unknown at the radar receiver. As expected, radar performance improves as the communications SINR threshold $\gamma_u$ decreases, since a lower $\gamma_u$ relaxes the communications constraints and enlarges the feasible design space.
For $L=1$, the performance without knowledge of the transmitted symbols is close to the known-symbol case. However, for $L>1$, knowledge of the transmitted signals enables STAP-based coherent processing across multiple observations, improving target detection and noise averaging. In contrast, when the signals are unknown, the observations lack exploitable structure, limiting the benefits of increasing $L$ and resulting in nearly constant performance. Overall, the proposed ISAC schemes approach the radar-only benchmark under moderate SINR requirements, demonstrating reliable joint radar--communications operation with minimal sensing degradation.

% Fig.~\ref{fig:Bistatic_L} illustrates the radar–communications performance for both scenarios, where the transmitted signals are either known or unknown at the radar receiver, for different frame lengths $L$. Due to the inherent tradeoff between radar and communications, radar performance improves as the communications SINR threshold $\gamma_u$ decreases. Lowering $\gamma_u$ relaxes the communications constraint, allowing the radar objective to improve.
% For $L=1$, the performance without knowledge of the transmitted symbols is close to that of the known-symbol case. However, when $L>1$, the receiver with known signals can exploit multiple observations to implement STAP filters, enhancing target detection. A longer frame allows the radar to average out noise across multiple samples, making knowledge of the transmitted symbols beneficial for coherent processing. In contrast, when the transmitted signals are unknown, the received observations are unstructured, limiting multi-time-slot gains; the performance remains nearly constant regardless of $L$. Overall, the proposed ISAC schemes achieve performance close to the radar-only benchmark under moderate communications SINR requirements, indicating that reliable joint radar–communications operation can be maintained with minimal sensing degradation.

% Fig.~\ref{fig:Bistatic_U} shows the effect of the number of communication users $U$. As $U$ increases, the achievable communications performance decreases because more users impose stricter SINR constraints, reducing the feasible design space for a given $\gamma_u$. 
Fig.~\ref{fig:Bistatic_U} shows the effect of the number of communication users $U$. Increasing $U$ reduces the achievable communication performance, as more users impose stricter SINR constraints and shrink the feasible design space for a given $\gamma_u$.
% The radar-only design corresponds to the special case $U=0$, where the optimization focuses exclusively on radar performance without communications constraints.

\section{Conclusions}\label{sec:conclusion}
In this paper, we studied joint beamforming and radar receive filter design in a multiuser bistatic ISAC system under communications SINR and power constraints. We developed algorithms for both known and unknown transmit-symbol cases at the radar receiver, using FP–SCA for the known case and an AO-based approach for the unknown case. Numerical results show that symbol knowledge yields additional gains in multi-slot processing, while single-slot performance remains comparable even without it. Moreover, the proposed ISAC designs perform close to the radar-only benchmark under moderate communications SINR requirements, confirming the effectiveness of the proposed algorithms.

\section{Acknowledgment}
This work has been financially supported in part by Infotech Oulu, the Research Council of Finland through the 6G Flagship Programme (grant number: 346208) and through project DIRECTION (grant number: 354901), and CHIST-ERA via project PASSIONATE (grant number: 359817). The research has also been supported in part by Business Finland, Digita, Patria, Saab, Bittium, and Nokia via the 6G-ISAC project, as well as Business Finland, Nokia, Keysight, and Finwe via the 6G-WISECOM project.

\bibliographystyle{IEEEtran}
\bibliography{Bib/conf_short,Bib/IEEEabrv,Bib/Bibliography}

\end{document}